\documentclass[aps,reprint,amsmath,amssymb]{revtex4-1}
\usepackage{graphicx}
\usepackage{dcolumn}
\usepackage{bm}
\begin{document}
\title{Strong coupling regime in coherent electron transport in periodic quantum nanostructures}
\author{L. S. Petrosyan$^{1,2,3}$ and T. V. Shahbazyan$^1$} 
\affiliation{
$^1$Department of Physics, Jackson State University, Jackson, Mississippi
39217 USA\\
$^2$Institute for Mathematics and High Technology, Russian-Armenian State University, 123 Hovsep Emin Street, Yerevan  0051, Armenia\\
$^3$Department of Medical Physics, Yerevan State Medical University, 2 Koryun Street, Yerevan 
0025, Armenia
}

\begin{abstract}
We study coherent transport in a system of a periodic linear chain of quantum dots placed between two parallel quantum wires. We show that  resonant-tunneling conductance between the wires exhibits a Rabi splitting of the resonance peak as a function of Fermi energy in the wires indicating the emergence of strong coupling between the system constituents. The underlying mechanism of the strong coupling regime is conservation of the  quasimomentum in a periodic system that leads to transition resonances between electron states in a quantum dot chain and quantum wires. A perpendicular magnetic field, by breaking the system's  left-right symmetry, gives rise to  a fine structure of the conductance lineshape.
\end{abstract}
\maketitle

\section{Introduction}
\label{intro}

During the past decade, strong coupling effects in the optics of nanostructures have been a subject of intense interest \cite{barnes-rpp15}. Optical interactions between excited dye molecules or excitons in semiconductor structures and resonant optical cavity modes or surface plasmons in metal structures can lead to a mixed state  with dispersion characterized by an anticrossing gap (Rabi splitting) in the resonance region. A strong coupling regime is established when the \textit{coherent} energy exchange between two systems exceeds  incoherent losses through radiative or nonradiative mechanisms, while the Rabi splitting magnitude can vary over a wide range \cite{forchel-nature04,khitrova-nphys06,imamoglu-nature07,bellessa-prl04,sugawara-prl06,wurtz-nl07,fofang-nl08,bellessa-prb09,fofang-nl11,guebrou-prl12,schlather-nl13,hakala-prl09,berrier-acsnano11,salomon-prl12,garcia-prl13,antosiewicz-acsphotonics14,luca-apl14,garcia-prl14,gomez-nl10,gomez-jpcb13,manjavacas-nl11,vasa-prl08,lawrie-nl12,velizhanin-prb14}. For example, a relatively weak Rabi splitting in the range $100$ $\mu$eV -- 1 meV was reported for a semiconductor quantum dot (QD) radiatively coupled to a cavity mode  \cite{forchel-nature04,khitrova-nphys06,imamoglu-nature07}, whereas a much larger splitting (above 100 meV) was observed for surface plasmons coupled  to excitons in J-aggregates \cite{bellessa-prl04,sugawara-prl06,wurtz-nl07,fofang-nl08,bellessa-prb09,fofang-nl11,guebrou-prl12,schlather-nl13}, individual dye molecules \cite{hakala-prl09,berrier-acsnano11,salomon-prl12,garcia-prl13,antosiewicz-acsphotonics14,luca-apl14,garcia-prl14}, or semiconductor QDs \cite{gomez-nl10,gomez-jpcb13,manjavacas-nl11}. A weaker, although still significant, Rabi splitting ($\sim$10 meV) was reported for quantum well excitons coupled to surface-plasmon polaritons \cite{vasa-prl08,lawrie-nl12} or to graphene plasmons \cite{velizhanin-prb14}. Recently, strong coupling between molecular vibrational modes and cavity modes in Raman scattering experiments was reported \cite{ebbesen-nc15, long-acsp15,ebbesen-ac15,garcia-njp15}.
\begin{figure}[tb]
\begin{center}
\includegraphics[width=0.99\columnwidth]{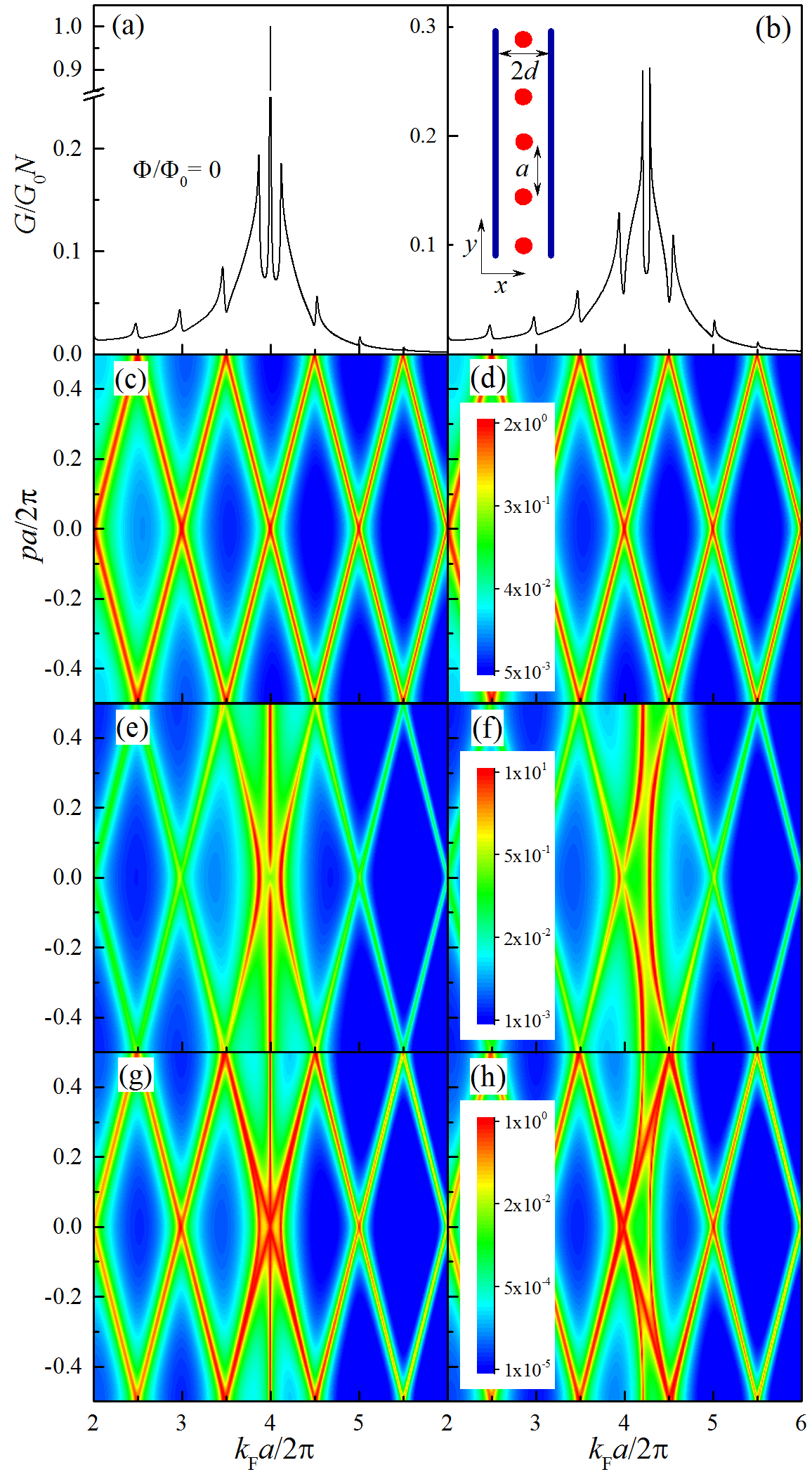}
\caption{\label{fig:1}
Normalized per QD conductance vs. Fermi momentum $k_{F}$ for resonance positions at (a) $k_{F}/k_{a}=4.0$ and (b)  $k_{F}/k_{a}=4.25$ is shown along with the respective density plots in the $(p,k_{F})$ plane of joint tunneling rate $\Gamma_{p}^{t}$ in units of $\Gamma_{a}$ [(c) and (d)], QDC spectral function $A_{p}$ in units of $\Gamma_{a}^{-1}$ [(e) and (f)], and partial conductivity $G_{p}$ [(g) and (h)]. The same color scale is used for each pair. Inset: QW/QDC/QW system schematics.}
\end{center}
\end{figure}

On the other hand, the  electron quantum transport in semiconductor nanostructures \cite{yacoby-prl95,shuster-nature97,buks-nature98} bears deep similarities to coherent optical processes \cite{brandes-pr05}. The  interference of electron pathways in confined structures gives rise to, e.g., the analog of Dicke superradiance in resonant tunneling  through several QDs \cite{shahbazyan-prb94,brandes-pr05}, Fabry-P\'{e}rot interference in electron waveguides \cite{park-nature01}, or  extraordinary electron transmission through a QD lattice \cite{petrosyan-prl11}. However, to the best of our knowledge, a physical mechanism that could give rise to a  strong coupling regime in electron transport  has yet to be suggested.
 
Here we demonstrate that a strong coupling regime can be realized in resonant tunneling through a  periodic chain of QDs (QDC) \cite{bandyopadhyay-nano96,liang-apl04} placed between two parallel semiconductor quantum wires (QW) \cite{yacoby-science02,yacoby-prl02} (see the inset in Fig. \ref{fig:1}).  Due to scattering by  the QDC periodic potential, the  energy spectrum of one-dimensional electron gas (1DEG) in QWs splits into Bloch bands that are characterized by a quasimomentum that conserves across the system. We show that even for weak tunneling between individual QDs and QWs, the momentum-selective transitions between the QDC states and Bloch states in QWs cause an anticrossing of the Bloch band  dispersion and resonant energy level, leading to Rabi splitting of the conduction peak. A perpendicular magnetic field, by breaking the symmetry between the left and right QWs, leads to a fine structure of the conductance lineshape.


\section{Conductance through a periodic array of quantum dots}
\label{theory}

We consider resonant tunneling through a QDC with lattice constant $a$  separated from the left and right QWs by potential barriers (see  Fig. \ref{fig:1}). Within tunneling Hamiltonian formalism \cite{brandes-pr05}, the  system Hamiltonian is
\begin{equation}
\label{h}
H=\sum_{j} E_{0}c_{j}^{\dagger}c_{j}+ 
\sum_{k\alpha}{\cal E}_{k}^{\alpha}c_{k\alpha}^{\dagger}c_{k\alpha}+
\sum_{k\alpha j} \left (V_{jk}^{\alpha} c_{j}^{\dagger}c_{k\alpha}+\text{H.c.}\right ),
\nonumber
\end{equation}
where  $ E_{0}$ and $c_{j}^{\dagger}$ ($c_{j}$) are, respectively, the energy and  creation (annihilation) operators for QD states, ${\cal E}_{k}^{\alpha}$ and $c_{k\alpha}^{\dagger}$ ($c_{k\alpha}$) are those for QW states with momentum $k$ ($\alpha=L,R$ stands for the left/right QW), and $V_{k j}^{\alpha}$ is the transition matrix element between the QD and QW states. We assume no direct tunneling between   QDs, and restrict ourselves to the single-electron picture of transport due to a low probability of QD double occupancy in a long chain. The zero-temperature conductance through $N$ QDs is \cite{brandes-pr05}
\begin{equation}
\label{cond}
G=\frac{e^{2}}{\pi\hbar}\, {\rm Tr} \left(\hat{\Gamma}^{R}\frac{1}{E_{F}-E_{0}-\hat{\Sigma}} \,\hat{\Gamma}^{L}\frac{1}{E_{F}-E_{0}-\hat{\Sigma}^{\dagger}}\right),
\end{equation}
where $\Sigma_{ij}=\Sigma_{ij}^{L}+\Sigma_{ij}^{R}$ is  the self-energy matrix of  QDC states due to the coupling to the left and right QWs, 
\begin{equation}
\label{self-matrix}
\Sigma_{ij}^{\alpha} =\sum_{ k}\frac{V_{ik}^{\alpha}V_{k j}^{\alpha}}{E_{F} -{\cal E}_{k}^{\alpha}+i\gamma_{\alpha}} = \Delta_{ij}^{\alpha}-\frac{i}{2} \Gamma_{ij}^{\alpha}.
\end{equation}
Here the real and imaginary parts of $\Sigma_{ij}^{\alpha}$ define the energy matrix $\Delta_{ij}^{\alpha}$ and the decay matrix $\Gamma_{ij}^{\alpha}$ and the trace is taken over $N$ QDC sites $y_{j}$. The matrix element can be presented as  $V_{jk}^{\alpha}=L^{-1/2}e^{iky_{j}}t_{\alpha}$, where $t_{\alpha}$ is  the tunneling amplitude between QDs and QWs, and $L=Na$ is the normalization length \cite{shahbazyan-prb94,brandes-pr05}. Then, the self-energy (\ref{self-matrix}) takes the form $\Sigma_{ij}^{\alpha} =t_{\alpha}^{2}G_{\alpha}(y_{i}-y_{j})$, where $G_{\alpha}(y_{i}-y_{j})$ is the electron Green's function in QWs.

Due to  QDC periodicity,  tunneling between QDC and QWs gives rise to a quasimomentum $p$ along the $y$ direction that conserves across the QW/QDC/QW system \cite{petrosyan-prl11,petrosyan-prb15}. The 1DEG  momentum space splits into Bloch bands, $k\rightarrow g_{n}+p$, where  $g_{n}=k_{a}n$ is the $n$th Bloch band wave vector ($k_{a}=2\pi/a$ is the reciprocal lattice vector, and $n$ is an integer). The QDC spectrum is derived through the Fourier transform of the self-energy matrix  as  $\Sigma_{ij}^{\alpha}=N^{-1}\sum_{p}e^{ip(y_{i}-y_{j})}\Sigma_{p}^{\alpha}$, where 
\begin{equation}
\label{self-lsb}
\Sigma_{p}^{\alpha}=\frac{t_{\alpha}^{2}}{a}\sum_{n}G^{\alpha}_{g_{n}+p} =\frac{t_{\alpha}^{2}}{a}\sum_{n} \frac{1}{E_{F} -{\cal E}_{g_{n}+p}^{\alpha}+i\gamma_{\alpha}} 
\end{equation}
is QDC self-energy in the momentum space. Here $G^{\alpha}_{g_{n}+p}$ is  the QW Green's function for a band $n$  electron with dispersion ${\cal E}_{g_{n}+p}^{\alpha}=\hbar^{2}\left (g_{n}+p\right )^{2}/2m_{\alpha}$ and scattering rate $\gamma_{\alpha}$  ($m_{\alpha}$ is the electron mass).  The real and imaginary parts of self-energy $\Sigma_{p}^{\alpha}=\Delta_{p}^{\alpha}-\frac{i}{2}\Gamma_{p}^{\alpha}$ determine, respectively, the QDC states' dispersion, $E_{p}=E_{0}+\Delta_{p}^{L}+\Delta_{p}^{R}$, and decay rate, $\Gamma_{p}=\Gamma_{p}^{L}+\Gamma_{p}^{R}$.

A perpendicular magnetic field ${\bm B}$,  included via the vector potential  ${\bm A}=(0, Bx,0)$, leads to the momentum shift  $k\rightarrow k \pm (e/\hbar c) dB$ in the left and right QWs  located at $x=\mp d$, respectively. In the presence of a periodic QDC potential, the Bloch bands in the left and right QWs are shifted in   \textit{opposite} directions, $k\rightarrow g_{n}^{L,R}(B)+p$, where
\begin{equation}
\label{band}
g_{n}^{L,R}(B)=k_{a}\left (n\pm \frac{\Phi}{2\Phi_{0}}\right )
\end{equation}
is the Bloch wave vector shifted by a magnetic flux  $\Phi=2daB$,   in units of the flux quantum $\Phi_{0}=hc/e$, through the elementary area enclosed by tunneling electron  paths. Here, we  assumed that the magnetic field does not cause any significant Zeeman splitting. 

Finally,  by performing the Fourier transform of Eq.~(\ref{cond}), the conductance can be expressed in terms of the system eigenstates as $G=N G_{0} k_{a}^{-1}\int \! d p \,G_{p}$, where 
\begin{equation}
\label{cond2}
G_{p}=\Gamma_{p}^{L}S_{p}\Gamma_{p}^{R}S_{p}^{\dagger}=A_{p} \Gamma_{p}^{t}
\end{equation}
is the dimensionless partial conductance (transmission coefficient) of the state $p$,   $S_{p}=\left (E_{F}-E_{p}-i\Gamma_{p}/2\right )^{-1}$ is the QDC Green's function (for brevity, $G_{0}=\pi e^{2}/\hbar$), and $p$ integration is taken over the Brillouin zone $(-\pi/a, \pi/a$). To elucidate  different contributions to $G_{p}$, we introduced the QDC spectral function, $A_{p}=-2\,{\rm Im} \, S_{p}$, and the joint tunneling rate associated with tunneling time across the system, $\Gamma_{p}^{t}=\left (1/\Gamma_{p}^{L}+1/\Gamma_{p}^{R}\right )^{-1}$; since $\Gamma_{p}^{L,R}=-2{\rm Im}\,\Sigma_{p}^{L,R}$ is proportional to the left/right QW spectral function [see Eq. (\ref{self-lsb})],  $\Gamma_{p}^{t}$ is determined by their \textit{overlap}.

\section{Numerical results and discussion}
\label{num}

Below we present the results of numerical calculations for a symmetric configuration, i.e., QDC at the midpoint between identical QWs ($\gamma_{\alpha}=\gamma$, $m_{\alpha}=m$, $t_{\alpha}=t$). The QDC period $a$ was chosen to set $E_{0}\approx 16E_{a}$, where $E_{a}=\hbar^{2}k_{a}^{2}/2m$ is a  geometric energy scale associated with QDC, so that the transmission resonance at $E_{F}=E_{0}$ occurs at the Fermi wave vector values  $k_{F}\approx 4k_{a}$ (two values, $k_{F}/k_{a} = 4.0$ and  4.25, were used). The electron scattering rate  $\gamma=\hbar v_{F}/l$, where $v_{F}$ is the Fermi velocity and  $l$ is the scattering length, was varied in the range from $\gamma=0.006E_{0}$ to $0.07E_{0}$ (or $\gamma/E_{a}$ in the range $0.1-1.1$), yielding $l/a$ in the range from 1 to 10; for  $a\sim 100$ nm, this corresponds to a low-to-intermediate mobility in the range  $10^{4}$--$10^{6}$ cm$^{2}$/Vs. We assume that the tunneling rate between individual QDs and QWs, $\Gamma=2mt^{2}/\hbar^{2}k_{F}$ \cite{brandes-pr05},  is  small and set $\Gamma/E_{0}=0.01$. We use the energy-independent rate $\Gamma_{a}=2mt^{2}/\hbar^{2}k_{a}=(k_{F}/k_{a})\Gamma $ to describe the tunnel coupling between  QDC and QWs.

In Fig.\ \ref{fig:1}, we show the zero-field conductance along with density plots of $\Gamma_{p}^{t}$, $A_{p}$,  and $G_{p}$ in the $(p,k_{F})$ plane for a resonance position at $k_{F}/k_{a} = 4.0$ (left column)  and $k_{F}/k_{a} = 4.25$ (right column). To highlight the role of  Bloch bands, all curves are plotted against $k_{F}$ (in units of $k_{a}$) rather than $E_{F}$. In the absence of magnetic field, the energy spectra in the left and right QWs coincide  and the joint tunneling rate  $\Gamma_{p}^{t}$ traces Bloch bands in the $(k_{F},p)$ plane for each QW [see Figs.\ \ref{fig:1}(c) or \ref{fig:1}(d)]. The QDC spectral function $A_{p}$ [Figs.\ \ref{fig:1}(e) and  \ref{fig:1}(f)] peaks at resonance energy $E_{0}$ (vertical lines at $k_{F}/k_{a} = 4.0$ and  4.25, respectively)  and also shows a periodic Bloch pattern due to tunnel coupling between the QDC and QW states. The striking feature is a pronounced \textit{anticrossing} in regions of the $(p,k_{F})$ plane where, e.g., the $n$th Bloch band dispersion $E_{F}=\hbar^{2}(g_{n}+p)^{2}/2m$ meets the resonance at $E_{0}$. Note that for $E_{0}$ coinciding with the Bloch band center ($p=0$), two anticrossings from  the $n$th and $(-n)$th Bloch bands are superimposed  [Fig.\ \ref{fig:1}(e)], while for general $E_{0}$ anticrossings occur at two different $p$  [Fig.\ \ref{fig:1}(f)]. The partial conductance $G_{p}=A_{p}\Gamma_{p}^{t}$, shown in Figs.\ \ref{fig:1}(g) and \ref{fig:1}(h), is determined by the overlap of available states in QDC and QWs and, in fact,  represents the \textit{map  of conducting states} in the $(k_{F},p)$ plane. The  conductance is obtained by $p$ integration of $G_{p}$ over the Brillouin zone and shows pronounced spikes at the Brillouin zone's center ($p=0$) and edges ($p=\pm \pi/a$) , i.e., at $k_{F}=\pi n/a$ [Figs.\ \ref{fig:1}(a) and   \ref{fig:1}(b)].  However, the standout feature is two prominent peaks near the resonance  with a peak-to-peak separation equal to the anticrossing gap in $A_{p}$. 

To estimate the anticrossing gap (Rabi splitting), e.g., for $p=0$ [Fig.\ \ref{fig:1}(a)], we note that the main contribution to the electron self-energy (\ref{self-lsb}) comes from two terms in the sum corresponding to $g_{n}=\pm nk_{a}$. Keeping just these terms, the resonances in $A_{p}$ (for $p=0$) are found to occur at $E_{F}=n^{2}E_{a}=E_{0}$ and $E_{F}=n^{2}E_{a}\pm \Delta_{R}$, where 
\begin{equation}
\label{rabi}
\Delta_{R}=\sqrt{\frac{2}{\pi}E_{a}\Gamma_{a}-\gamma^{2}}
\end{equation}
is the Rabi splitting, which indicates the onset of a strong coupling regime. Note that when plotted against $k_{F}$, the Rabi splitting is $k_{F}\Delta_{R}/2E_{F}$. 

%
\begin{figure}[tb]
\begin{center}
\includegraphics[width=1.0\columnwidth]{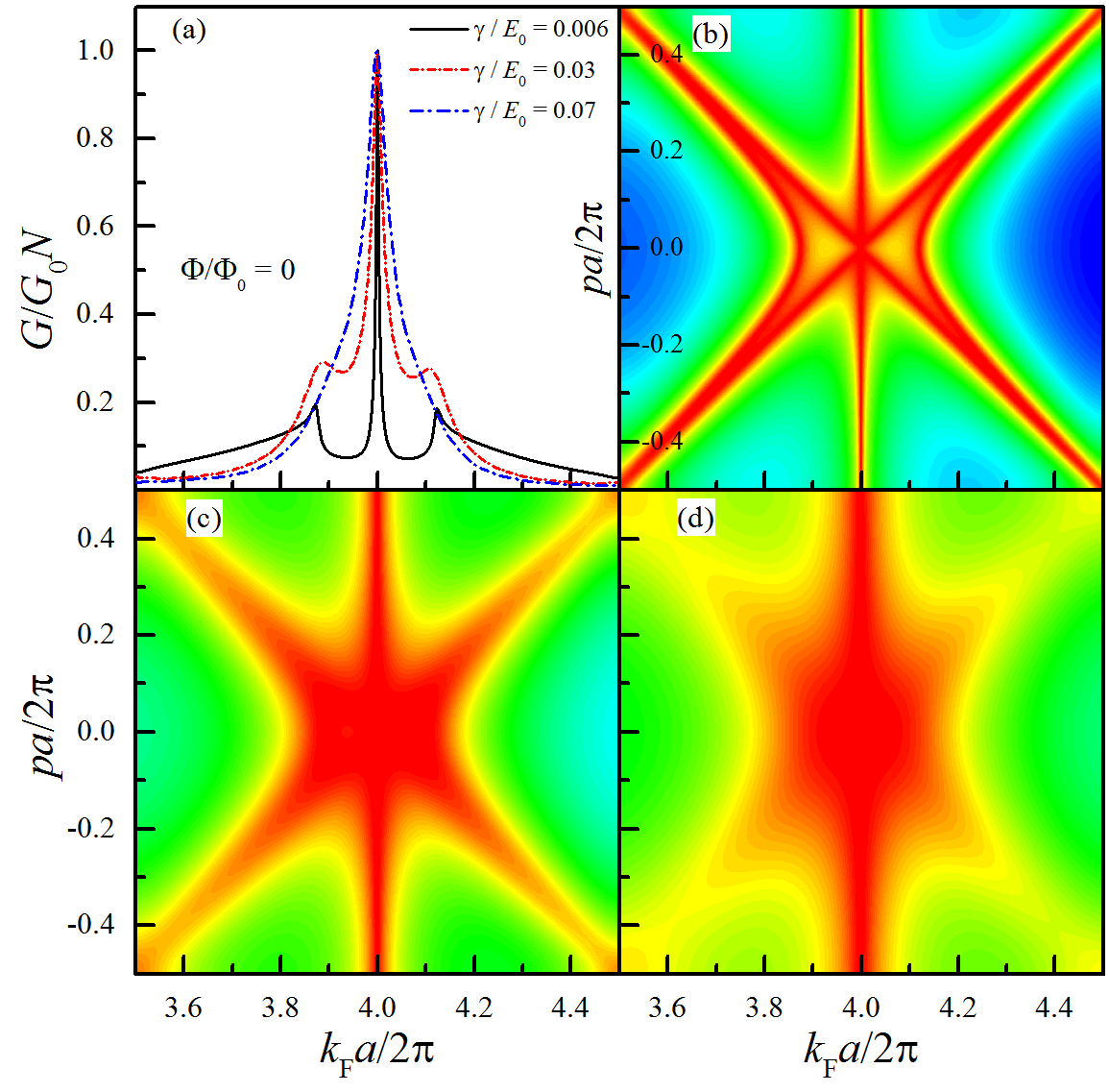}
\caption{\label{fig:2}
(a) Normalized per QD conductance is shown with increasing $\gamma$.  (b)--(d) show density plots of $G_{p}$ for $\gamma/E_{0}=0.007$,  0.03, and  0.07, respectively. The same  color scale for $G_{p}$ as in Fig. \ref{fig:1} is used.
}
\end{center}
\end{figure}
%

The emergence of a strong coupling regime in a periodic system with   weak tunneling can be traced to the  conservation of quasimomentum $p$ that leads to transition resonances between the QDC and QWs at the Bloch bands' energy shell [see Eq. (\ref{self-lsb})]. Note that this momentum-selective mechanism is somewhat analogous to resonant coupling between quantum well excitons and plasmons on a metal surface \cite{vasa-prl08} or in graphene \cite{velizhanin-prb14}, where in-plane momentum conservation  by the Coulomb coupling leads to the anticrossing of exciton and plasmon dispersions in the momentum space. The magnitude of Rabi splitting  (\ref{rabi}) depends on  the tunnel coupling $\Gamma_{a}$ between  QDC and QWs and on the electron mobility in QWs characterized by the scattering rate $\gamma$.  The latter parameter determines the \textit{effective system periodicity}, i.e., the number of QDs  in the chain, $N_{c}=l/a$,   the tunneling electron can visit before losing coherence. From Eq. (\ref{rabi}),  the onset of a strong coupling regime (i.e., $\Delta_{R}=0$) can be expressed as $N_{c} \sim \sqrt{E_{F}/\Gamma_{a}}$, indicating that smaller tunnel coupling between QDC and QWs requires a larger electron mobility  to reach the onset. In Fig. \ref{fig:2}, we show the conductance peak evolution with changing $\gamma$ along with the corresponding partial conductances.  With increasing $\gamma$, the  anticrossing  gap and hence the conductance peak Rabi splitting gradually diminish before  eventually disappearing  for $\gamma$ exceeding the critical value.

\begin{figure}[tb]
\begin{center}
\includegraphics[width=0.99\columnwidth]{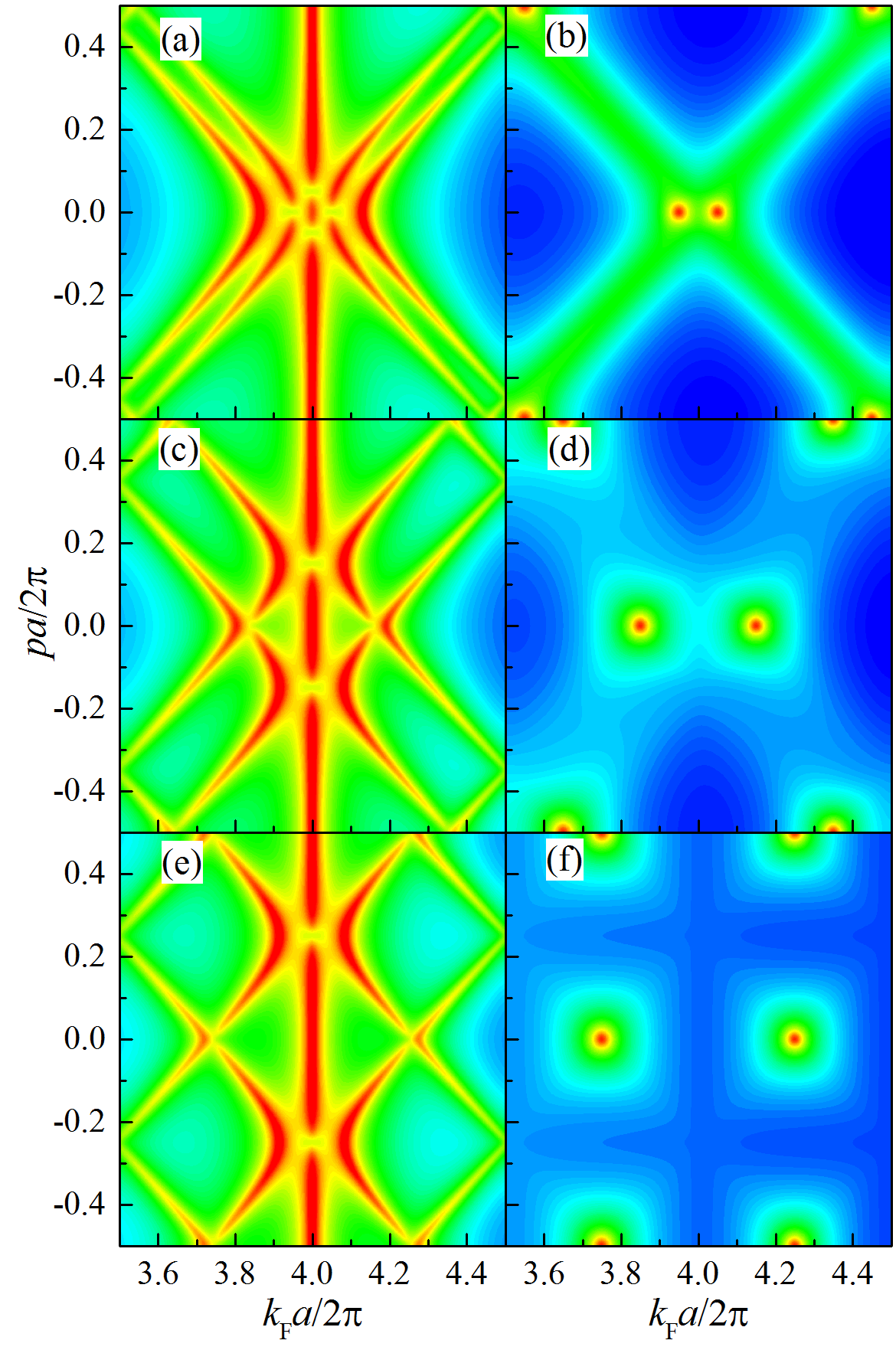}
\caption{\label{fig:3}
Density plots of $A_{p}$ (left column) and $\Gamma_{p}^{t}$ (right column) are shown with increasing magnetic field for $\Phi/\Phi_{0}=0.1$ [(a) and (b)], $\Phi/\Phi_{0}=0.3$ [(c) and (d)], and $\Phi/\Phi_{0}=0.5$ [(e) and (f)]. The same  color scales and units for $A_{p}$ and $\Gamma_{p}^{t}$ as in Fig. \ref{fig:1} are used.
}
\end{center}
\end{figure}
\begin{figure}[tb]
\begin{center}
\includegraphics[width=0.99\columnwidth]{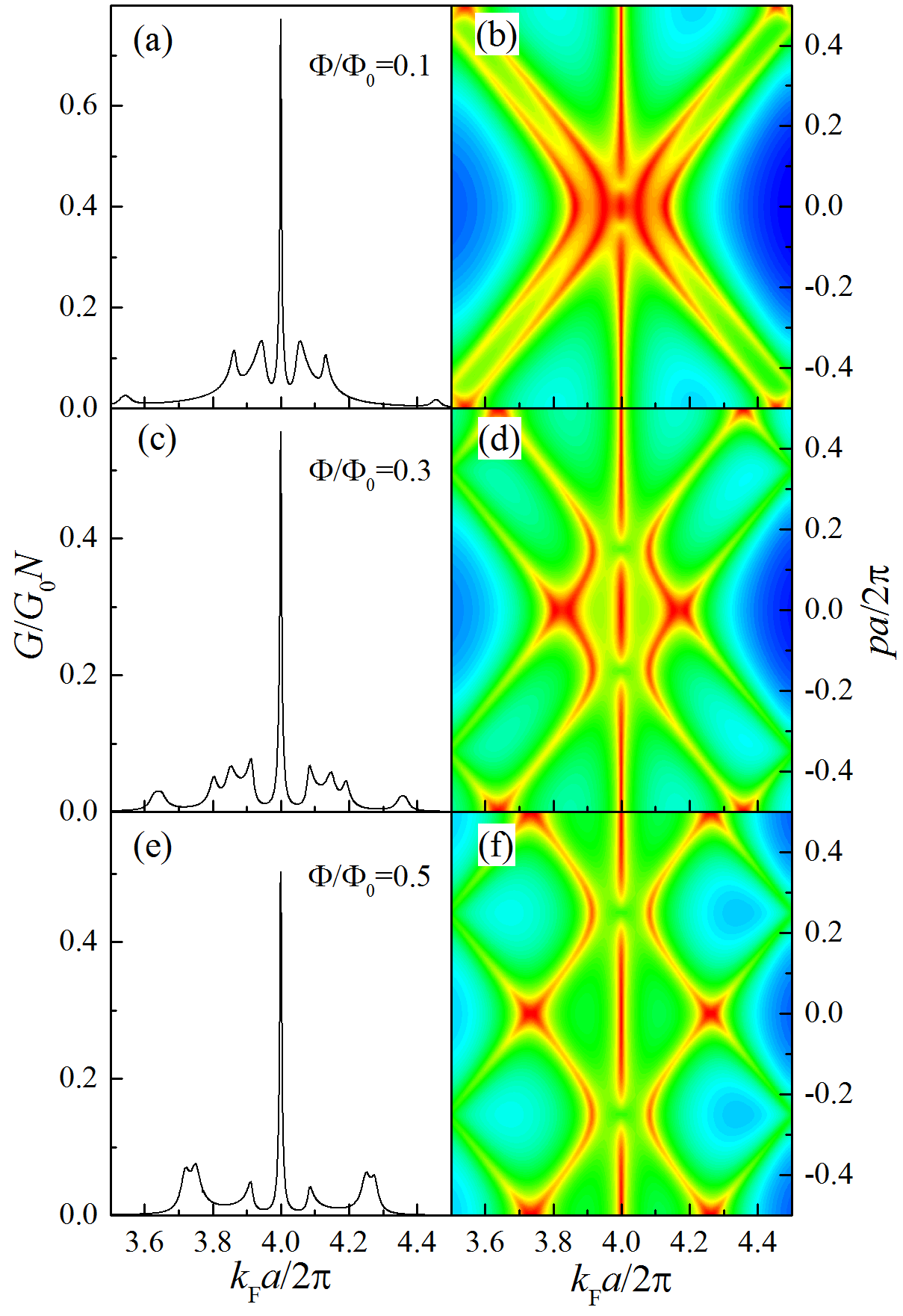}
\caption{\label{fig:4}
Normalized per QD conductance for resonance position at $k_{F}/k_{a}=4.0$ (left column) is shown with increasing magnetic field along with the corresponding density plot of $G_{p}$ (right column). The same  color scale for $G_{p}$ as in Fig. \ref{fig:1} is used.
}
\end{center}
\end{figure}

Turning  on the magnetic field shifts the left and right QW Bloch bands in the opposite directions along the $p$ axis (see Fig. \ref{fig:3}), leading to several anticrossings at different values of $p$, as shown in the $A_{p}$ density plot (left column). At the same time, the overlap between the QW spectral functions, characterized by $\Gamma_{p}^{t}$, is reduced to several intersection points between the left and right QW Bloch band sets (right column). With increasing field, the anticrossings in $A_{p}$ move away from each other along the $p$ axis, following the Bloch bands' field dependence [Figs.\ \ref{fig:3}(a), \ref{fig:3}(c), and \ref{fig:3}(e)], while the  maxima of $\Gamma_{p}^{t}$ move away from each other along the $k_{F}$ axis [panels (b), (d), and (f)], tracking the intersection points. This results in a fine field-induced structure of the conductance  lineshape, shown in Fig. \ref{fig:4} along with the density plot of  partial conductance $G_{p}=A_{p}\Gamma_{p}^{t}$. With increasing field, the conductance develops multiple peaks as $k_{F}$ changes within the Brillouin zone [Figs.\ \ref{fig:4}(a), \ref{fig:4}(c), and \ref{fig:4}(e)] in accordance with the evolution of conducting states in the $(p,k_{F})$ plane [Figs.\ \ref{fig:4}(b), \ref{fig:4}(d), and \ref{fig:4}(f)]. At the same time, the overall conductance magnitude is reduced due to a redistribution of the oscillator strengths between multiple peaks.

\begin{figure}[tb]
\begin{center}
\includegraphics[width=0.99\columnwidth]{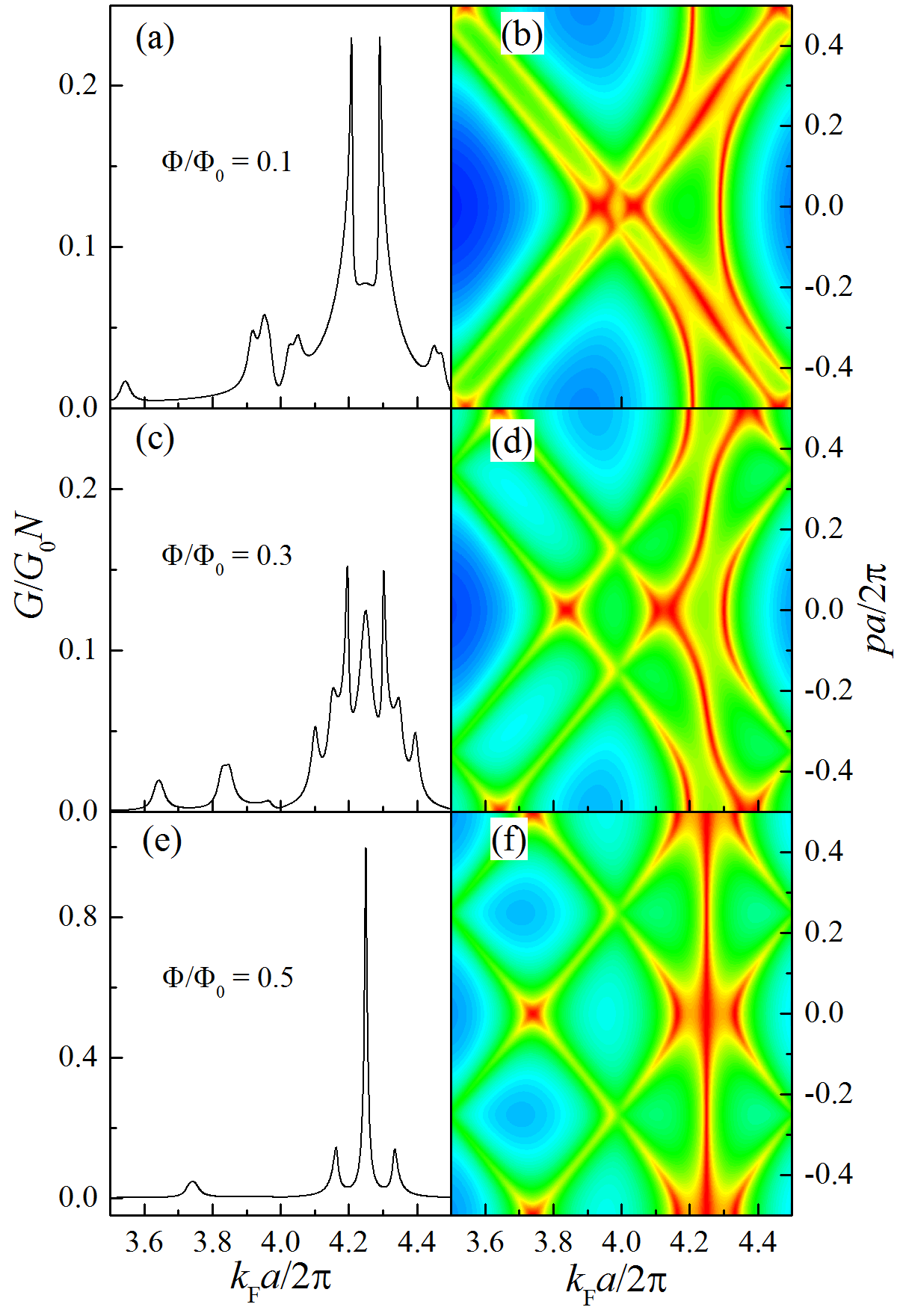}
\caption{\label{fig:5}
Normalized conductance for resonance position at $k_{F}/k_{a}=4.25$ (left column) is shown with increasing magnetic field along with the corresponding density plot of $G_{p}$ (right column). The same  color scale for $G_{p}$ as in Fig. \ref{fig:1} is used.
}
\end{center}
\end{figure}

By moving the resonance position $E_{0}$ away from the Bloch band center (e.g., to $k_{F}/k_{a}=4.25$,  as shown in Fig. \ref{fig:5}),  the central conductance peak disappears (compare to Fig. \ref{fig:1}), while the overall lineshape still exhibits multiple peaks. With increasing field, however, the resonance peak reappears [Figs.\ \ref{fig:5}(c) and \ref{fig:5}(d)] and, for $\Phi/\Phi_{0}=1/2$, it is nearly fully restored  while, at the same time, the fine structure largely disappears and only a single Rabi splitting remains intact [Figs.\ \ref{fig:5}(e) and \ref{fig:5}(f)]. This is related to the recovery of the left-right symmetry for magnetic field values satisfying $k_{F}=k_{a}|n\pm\Phi/2\Phi_{0}|$ [see Eq. (\ref{band})]; for such fields, both the left QW $n$th Bloch band and the right QW $(-n)$th Bloch band meet the resonance at the $p=0$ point in the $(p,k_{F})$ plane [\ref{fig:5}(f)]. Note, finally, that the conductance exhibits a usual Aharonov-Bohm periodicity for integer values of $\Phi/\Phi_{0}$ (not shown here).

\section{Conclusion}
\label{conc}

In summary, we have shown that   resonant tunneling conductance through a periodic chain of quantum dots placed between two parallel quantum wires can exhibit  Rabi splitting of the resonance peak as a function of Fermi energy due to strong coupling between the electron states in quantum dot chain and quantum wires. The underlying mechanism of strong coupling here is the conservation of quasimomentum $p$ that strongly enhances the transition amplitude between  quantum dot chain and quantum wires when Bloch band  dispersion in quantum wires approaches the quantum dot level.  This novel effect in coherent transport is analogous to the Rabi splitting in optical spectra of two interacting systems caused by the anticrossing gap in the energy spectrum of mixed state. A perpendicular magnetic field breaks the symmetry between left and right quantum wires leading to a fine structure of the conductance lineshape.

\acknowledgments

This work was supported by the National Science Foundation under Grant No. DMR-1206975. 
L.S.P. acknowledges support from the State Committee of Science, Republic of Armenia.

\end{document}